\documentclass[a4paper,11pt]{article}

\usepackage{latexsym}
\usepackage{amssymb}
\usepackage{amsfonts}
\usepackage{amsmath}
\usepackage{indentfirst}
\usepackage{xcolor}
\usepackage[dvips]{graphicx}
\usepackage{epsfig}

\newcommand{\cvd}{\hfill $\blacksquare$\bigskip}

\newtheorem{proposition}{Proposition}[section]

\newtheorem{corollary}{Corollary}[section]

\date{}

\author{S. Bilotta\thanks{Dipartimento di Sistemi e Informatica, viale
Morgagni 65, 50134 Firenze, Italy {\tt bilotta@dsi.unifi.it\quad
elisa@dsi.unifi.it\quad pinzani@dsi.unifi.it}}\and E. Pergola$^*$
 \and R. Pinzani$^*$\and S. Rinaldi\thanks{Dipartimento di Scienze
Matematiche ed Informatiche, Pian dei Mantellini, 44, 53100,
Siena, Italy {\tt rinaldi@unisi.it}}}

\title{Recurrence relations versus succession rules}

\begin{document}

\maketitle

\begin{abstract}
In this paper we present a method to pass from a recurrence
relation having constant coefficients (in short, a $C$-recurrence)
to a finite succession rule defining the same number sequence. We
recall that succession rules are a recently studied tool for the
enumeration of combinatorial objects related to the ECO method. We
also discuss the applicability of our method as a test for the
positivity of a number sequence.
\end{abstract}

\textbf{keywords:} Positivity problem; recurrence relation;
succession rule.

\section{Introduction}

Succession rules (sometimes called {\em ECO-systems}) have been
proved an efficient tool to solve several combinatorial problems.
A \emph{succession rule} $\Omega$ is a system constituted by an
\emph{axiom} $(a)$, with $a \in \mathbb{N}$, and a set of
\emph{productions} of the form:
\begin{displaymath}
(k)\rightsquigarrow (e_1(k))(e_2(k))\ldots(e_k(k)), \ \ \ \ \ k
\in \mathbb{N}, \ e_i : \mathbb{N} \rightarrow \mathbb{N}.
\end{displaymath}

A production constructs, for any given label $(k)$, its
\emph{successors} $(e_1(k)),(e_2(k)),\ldots,(e_k(k))$. The rule
$\Omega$ can be represented by means of a \emph{generating tree}
having $(a)$ as the label of the root and each node labelled $(k)$
at level $n$ produces $k$ sons labelled
$(e_1(k)),(e_2(k)),\ldots,(e_k(k))$, respectively, at level $n+1$.

A succession rule $\Omega$ defines a sequence of positive integers
$\{f_n\}_{n \geq 0}$ where $f_n$ is the number of the nodes at
level $n$ in the generating tree defined by $\Omega$. By
convention the root is at level 0, so $f_0=1$. The function
$f_{\Omega}(x)=\sum_{n \geq 0} f_n x^n$ is the \emph{generating
function} determined by $\Omega$.

\smallskip

The concept of a succession rule was introduced in \cite{CGHK} by
Chung et al. to study reduced Baxter permutations, and was later
applied to the enumeration of permutations with forbidden
subsequences \cite{West}.

Only later this has been recognized as an extremely useful tool
for the ECO method, a methodology applied for the enumeration
\cite{Eco1}, random generation \cite{rangen}, or exhaustive
generation \cite{exhgen} of various combinatorial structures.

More recently, succession rules have been considered as a
remarkable object to be studied independently of their
applications, and they have been treated by several points of
view. In \cite{genfun}, Banderier et al. explore in detail the
relationship between the form and the generating function of a
succession rule, and then provide a classification of rules as
{\em rational}, {\em algebraic}, or {\em transcendental},
according to their generating function type; besides some
algebraic properties of succession rules -- represented in terms
of a  {\em rule operator} --  have been determined and studied in
\cite{oper}.

Furthermore, some extensions of the concept of succession rule
have been proposed. In \cite{FPPR} the authors admit that a label
produces sons at different levels of the generating tree,
introducing the so called {\em jumping succession rules}, while in
\cite{C} the author presents {\em marked succession rules}, where
labels can be marked or not. The main difference with {\em
ordinary succession rules} is that marked labels annihilate
non-marked labels having the same value and lying at the same
level.

Finally, to explore the relationship between succession rules and
other formal tools for the enumeration is a rather popular trend
of research. A tentative in this direction has been made in
\cite{DFR} with the definition of the so called \emph{production
matrix}, which is just a way to represent a given succession rule
in terms of an infinite matrix, and supplies the possibility to
work with succession rules using some of the algebraic tools
developed for matrices.

More recently, there have been some efforts in developing methods
to pass from a recurrence relation defining an integer sequence to
a succession rule defining the same sequence - in this case we say
that the succession rule and the recurrence relation are {\em
equivalent}.

Our work fits into this research line, and tries to deepen the
relations between succession rules and recurrence relations.

It is worth mentioning that almost all studies realized until now
on this topic have regarded linear recurrence relations with
integer coefficients  \cite{BR,DFPR}. Following Zeilberger
\cite{Z}, we will address to these as {\em C-finite recurrence
relations}, and to the defined sequences as {\em C-finite
sequences}.

Accordingly, our work will start considering C-finite recurrences.
Compared with the methods presented in  \cite{BR,DFPR}, our
approach is completely different.

To achieve this goal, we first translate the given C-finite
recurrence relation into an {\em extended succession rule}, which
differs from the {\em ordinary} succession rules since it admits
both jumps and marked labels. Then we recursively eliminate jumps
and marked labels from such an extended succession rule, thus
obtaining a finite succession rule equivalent to the previous one.
We need to point out that this translation is possible only if a
certain condition -- called {\em positivity condition} -- is
satisfied. Such a condition ensures that all the labels of the
generating tree are non marked, hence the sequence defined by the
succession rule has all positive terms.

If the recurrence relation has degree $k$ with coefficients $a_1,
\ldots ,a_k$, such a condition can be expressed in terms of a set
of $k$ inequalities which can be obtained from a set of quotients
and remainders given by the coefficients. To the authors'
knowledge, such a condition is completely new in literature. It
directly follows that our positive condition provides a sufficient
condition for testing the positivity of a C-finite sequence then
it relates to the so called {\em positivity problem}.\\

\noindent {\bf Positivity Problem:} given a C-finite sequence
$\{f_n \}_{n\geq 0} $, establish if all its terms are positive.

This problem was originally proposed as an open problem in
\cite{BM}, and then re-presented in \cite{SS} (Theorems 12.1-12.2,
pages 73-74), but no general solution has been found yet.

It is worth mentioning that the positivity problem can be solved
for a large class of C-finite sequences, precisely those whose
generating function is a $\Bbb{N}$-rational series. We also recall
that the class of $\Bbb{N}$-rational series is precisely the class
of the generating functions of regular languages, and that
Soittola's Theorem \cite{S}  states  that the problem of
establishing whether a rational generating function is
$\Bbb{N}$-rational is decidable.

Soittola's Theorem has recently been proved in different ways in
\cite{BR,P}, using different approaches  and some algorithms to
pass from an $\Bbb{N}$-rational series to a regular expression
enumerated by such a series have been proposed \cite{BDFR,K}.
However, none of these techniques provides a method to face
C-finite recurrence relations which are not $\Bbb N$-rational.

Following the attempt of enlightening some questions on positive
sequences, some researches have recently focused on determining
sufficient conditions to establish the eventual positivity of a
given C-finite recurrence relation, as interestingly described in
\cite{G}. As a matter of fact, up to now, we only know that the
positivity problem is decidable for C-finite recurrences of two
\cite{HHH} or three terms \cite{LT}. Another possible approach to
tackle the positivity problem is to develop algorithms to test
eventual positivity of recursively defined sequences (and, in
particular, C-finite sequences) by means computer algebra, as in
\cite{Ge}.

Our work fits into this research line, since the positive
condition we propose is a sufficient condition for testing the
positivity of a C-finite sequence.

In the last section we consider the more general case of holonomic
integer sequences, i.e., those satisfying a linear recurrence with
polynomial coefficients. We show that also in this case we can
easily translate the recurrence relation into an infinite
succession rule (possibly having marked labels and jumps). A
further goal is to find a way to convert such a rule into an
ordinary succession rule, and find a more general criterion for
proving the positivity of an holonomic sequence. Here we present
an example in which we show a possible strategy to perform the
desired conversion.

\section{Basic definitions and notations}

In this section we present some basic definitions and notations
related to the concept of succession rule. For further definitions
and examples we address the reader to \cite{CGHK}.

An \emph{(ordinary) succession rule} $\Omega$ can be written in
the compact notation:
\begin{equation}
\label{uno} \left\{
\begin{array}{cl}
 (a) & \\
 (k) & \rightsquigarrow (e_1(k))(e_2(k))\ldots(e_k(k)) \,
\end{array}
\right.
\end{equation}

The rule $\Omega$ can be represented by means of a
\emph{generating tree}, that is a rooted tree whose vertices are
the labels of $\Omega$; where $(a)$ is the label of the root and
each node labelled $(k)$ produces $k$ sons labelled
$(e_1(k)),(e_2(k)),\ldots,(e_k(k))$, respectively. As usual, the
root lies at level 0, and a node lies at level $n$ if its parent
lies at level $n-1$. If a succession rule describes the growth of
a class of combinatorial objects, then a given object can be coded
by the sequence of labels in the unique path from the root of the
generating tree to the object itself.


A succession rule $\Omega$ defines a sequence of positive integers
$\{f_n\}_{n \geq 0}$ where $f_n$ is the number of the nodes at
level $n$ in the generating tree defined by $\Omega$. By
convention the root is at level 0, so $f_0=1$. The function
$f_{\Omega}(x)=\sum_{n \geq 0} f_n x^n$ is the \emph{generating
function} determined by $\Omega$.

Two succession rules are \emph{equivalent} if they have the same
generating function.  A succession rule is \emph{finite} if it has
a finite number of labels and productions.

For example, the two succession rules:
$$
\left \{
\begin{array}{l}
(2) \\
(2) \rightsquigarrow (2)(2)
\end{array}
\right. \qquad \qquad \left \{
\begin{array}{l}
(2) \\
(k) \rightsquigarrow (1)^{k-1}(k+1)
\end{array}
\right.
$$
are equivalent rules, and define the sequence $f_n=2^n$. The one
on the left is a finite rule, since it uses only the label $(2)$,
while the one on the right is an infinite rule.

A slight generalization of the concept of ordinary succession rule
is provided by the so called \emph{jumping succession rule}.
Roughly speaking, the idea is to consider a set of succession
rules acting on the objects of a class and producing sons at
different levels.

The usual notation to indicate a jumping succession rule $\Omega$
is the following:
\begin{equation}
 \left\{
\begin{array}{cl}
 (a) & \\
 (k) & \stackrel{j_1}{\rightsquigarrow} (e_{11}(k))(e_{12}(k))\ldots(e_{1k}(k))\\
 (k) & \stackrel{j_2}{\rightsquigarrow}
 (e_{21}(k))(e_{22}(k))\ldots(e_{2k}(k))\\
 \vdots\\
(k) & \stackrel{j_m}{\rightsquigarrow}
(e_{m1}(k))(e_{m2}(k))\ldots(e_{mk}(k))
\end{array}
\right.
\end{equation}

The generating tree associated with $\Omega$ has the property that
each node labelled $(k)$ lying at level $n$ produces $k$ sets of
sons at level $n+j_1$, $n+j_2$, $\dots$, $n+j_m$, respectively and
each of such set has labels
$(e_{i1}(k)),(e_{i2}(k)),\ldots,(e_{ik}(k))$ respectively, $1 \leq
i \leq m$.

A jumping succession rule $\Omega$ defines a sequence of positive
integers $\{f_n\}_{n \geq 0}$ where, as usual, $f_n$ is the number
of the nodes at level $n$ in the generating tree of $\Omega$. The
function $f_{\Omega}(x)=\sum_{n \geq 0} f_n x^n$ is the
\emph{generating function} determined by $\Omega$.

We need to point out that, according to the above definitions, a
node labelled $(k)$ has precisely $k$ sons. A rule having this
property is said to be \emph{consistent}. However, in many cases
we can relax this constraint and consider rules (as in \cite{C}),
where the number of sons is a function of the label $k$.

\begin{figure}[htb]
\begin{center}
\epsfig{file=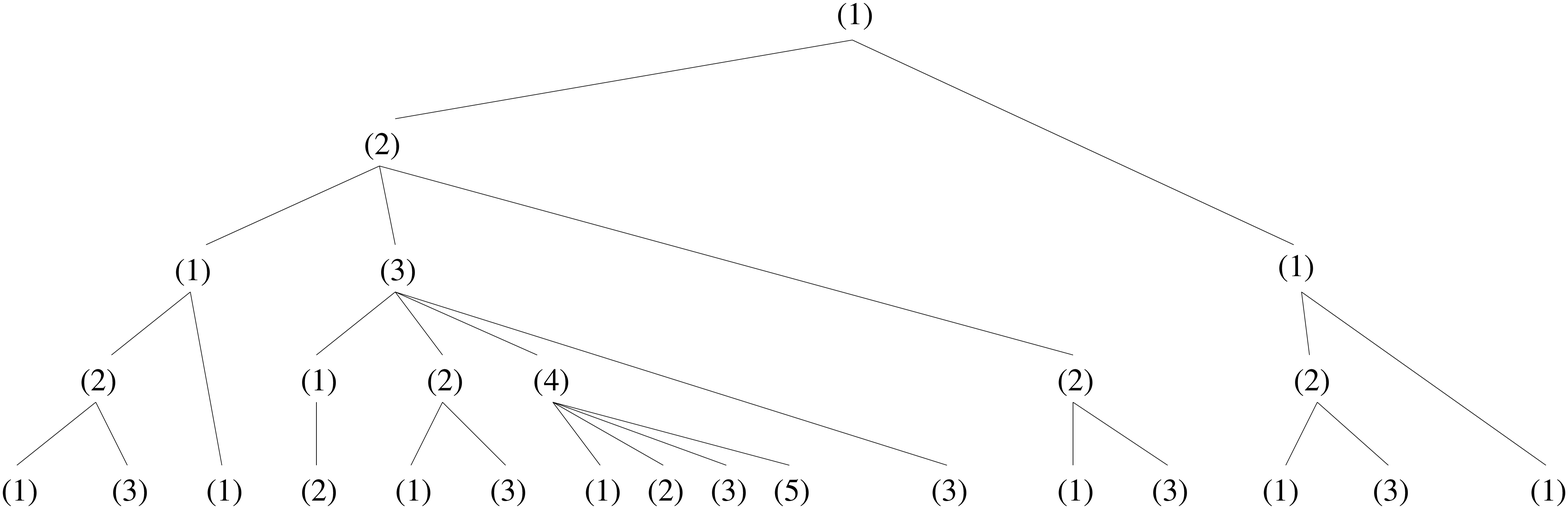,width=4.8in,clip=} \caption{\small{Four
levels of the generating tree associated with the succession rule
(\ref{jump})} \label{jp}}\vspace{-15pt}
\end{center}
\end{figure}

For example, the jumping succession rule (\ref{jump}) counts the
number of \emph{2-generalized Motzkin paths} and Figure \ref{jp}
shows some levels of the associated generating tree. For more
details about these topics, see \cite{FPPR}.
\begin{equation}\label{jump}
\left\{ \begin{array}{lll} (1)&
\\ (k)&\stackrel{1}{\rightsquigarrow} (1)(2)\cdots (k-1)(k+1)
\\ (k)&\stackrel{2}{\rightsquigarrow} (k)
\end{array}\right.
\end{equation}

Another generalization of succession rules is introduced in
\cite{MV}, where the authors deal with \emph{marked succession
rules}. In this case the labels appearing in a succession rule can
be marked or not, therefore \emph{marked} are considered together
with usual labels. In this way a generating tree can support
negative values if we consider a node labelled $(\overline{k})$ as
opposed to a node labelled $(k)$ lying on the same level.

A \emph{marked generating tree} is a rooted labelled tree where
 marked or non-marked labels appear according to the corresponding
succession rule. The main property is that, on the same level,
marked labels kill or annihilate the non-marked ones with the same
label value, in particular the enumeration of the combinatorial
objects in a class is the difference between the number of
non-marked and marked labels lying on a given level.

For any label $(k)$, we introduce the following notation for
generating tree specifications:
\begin{itemize}
\item[] $(\overline{\overline{k}})=(k)$; \item[] $(k)^n =
\underbrace{(k)\ldots(k)}_{n} \ \ n > 0;$ \item[] $(k)^{-n} =
\underbrace{(\overline{k})\ldots(\overline{k})}_{n} \ \ n > 0.$
\end{itemize}

\begin{figure}[htb]
\begin{center}
\epsfig{file=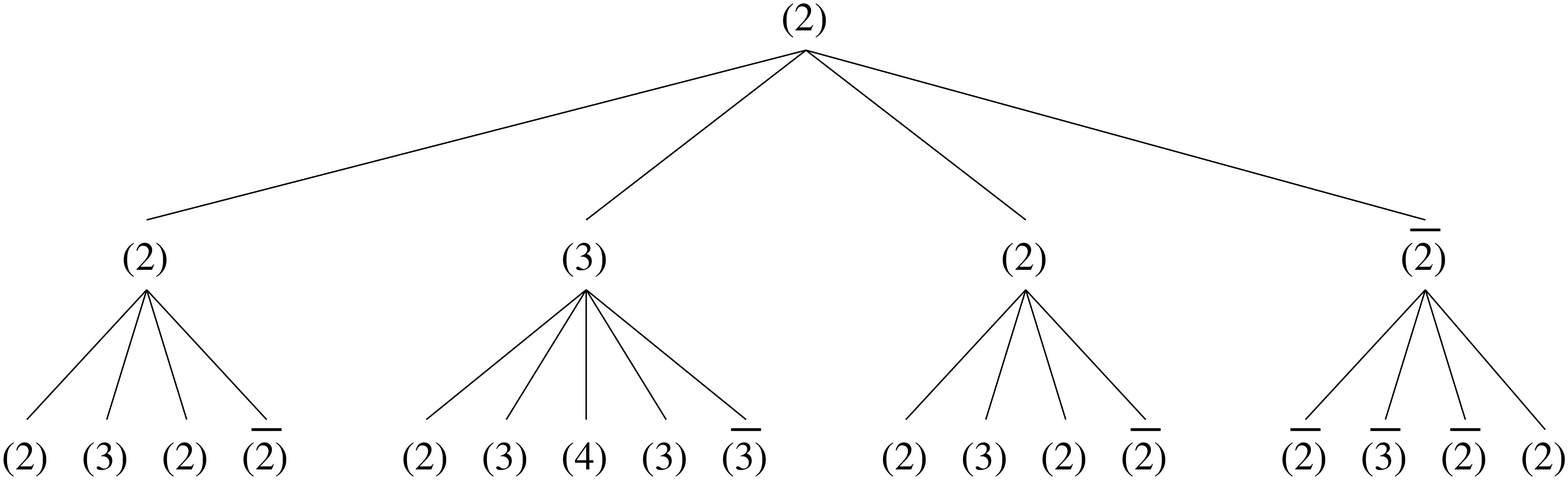,width=4.3in,clip=} \caption{\small{Three
levels of the generating tree associated with the succession rule
(\ref{quattro})} \label{exmar}}\vspace{-15pt}
\end{center}
\end{figure}

For example, the classical succession rule for Catalan numbers can
be rewritten in the form (\ref{quattro}) and Figure \ref{exmar}
shows some levels of the associated generating tree.
\begin{equation}
\label{quattro} \left\{
\begin{array}{ll}
 (2) & \\
 (k) & \stackrel{1}{\rightsquigarrow} (2)(3)\ldots(k)(k+1)(k)\\
 (k) & \stackrel{1}{\rightsquigarrow} (\overline{k})
\end{array}
\right.
\end{equation}

The concept of marked labels has been implicity used for the first
time in \cite{MSV}, then in \cite{C} in relation with the
introduction of signed ECO-systems.

\section{A general method to translate C-sequences into succession
rules}\label{method}

The main purpose of our research is to develop a general formal
method to translate a given recurrence relation into a succession
rule defining the same number sequence. By abuse of notation, in
this case we will say that the recurrence relation and the
succession rules are {\em equivalent}.

As a first step we deal with linear recurrence relations with
integer coefficients \cite{BR,DFPR}. Following Zeilberger
\cite{Z}, we will address to these as {\em C-finite recurrence
relations}, and to the defined sequences as {\em C-finite
sequences}.

\smallskip

\noindent This section is organized as follows.

\begin{description}
\item{i)} {First we deal with C-sequences of the form
\begin{equation}\label{fn}
f_n = a_1 f_{n-1} + a_2 f_{n-2} + \dots + a_k f_{n-k} \ \ \ \ \
a_i \in \mathbb{Z}, 1 \leq i \leq k
\end{equation}
with {\em default} initial conditions, i.e. $f_0=1$ and $f_h=0$
for all $h<0$. We translate the given C-finite recurrence relation
into an extended succession rule, possibly using both jumps and
marked labels (Section \ref{s1}).}

\item{ii)} {Then, we recursively eliminate jumps and marked labels
from such an extended succession rule, thus obtaining a finite
succession rule equivalent to the previous one (Section \ref{s2}).
We remark that steps i) and ii) can be applied independently of
the positivity of $\{ f_n \}_{n \geq 0}$, but at this step we
cannot be sure that all the labels of the obtained rule are
nonnegative integers.}

\item{iii)} {We state a condition to ensure that the labels of the
obtained succession rule are all nonnegative. If such a condition
holds, then the sequence $\{ f_n \}_{n \geq 0}$ has all positive
terms, thus we refer to this as {\em positivity condition}
(Section \ref{s3}).}

\item{iv)} {We show how our method can be extended to
$C$-sequences with generic initial conditions (Section \ref{s4}).}
\end{description}

\subsection{C-sequences with default initial conditions}\label{s1}

Let us consider a C-finite recurrence relation expressed as in
(\ref{fn}), with default initial conditions and the related
C-sequence $\{ f_n \} _{n \geq 0}$. We recall that the generating
function of $\{ f_n \}_{n \geq 0}$ is rational, and precisely it
is
\begin{equation} \label{fungen}
f(x)=\sum_{n \geq 0} f_n x^n = \frac{1}{1- a_1 x - a_2 x^2 - \dots
- a_k x^k} \, .
\end{equation}

The first step of our method consists into translating the
C-finite recurrence relation (\ref{fn}) into an extended
succession rule. The translation is rather straightforward, since
in practice it is just an equivalent way to represent the
recurrence relation.

\begin{proposition}
The recurrence relation (\ref{fn}) with default initial conditions
is equivalent to the following extended succession rule:
\begin{equation} \label{mark} \left\{
\begin{array}{cl}
 (a_1) & \\
 (a_1) & \stackrel{1}{\rightsquigarrow} (a_1)^{a_1}\\
 (a_1) & \stackrel{2}{\rightsquigarrow} (a_1)^{a_2}\\
  \vdots\\
 (a_1) & \stackrel{k}{\rightsquigarrow} (a_1)^{a_k}
\end{array}
\right.
\end{equation}
\end{proposition}

For example, the recurrence relation $f_n=
3f_{n-1}+2f_{n-2}-f_{n-3}$ with default initial conditions,
defines the sequence $1,3,11,38,133,464,1620,5655,  \ldots ,$ and
it is equivalent to the following extended succession rule:
\begin{equation} \label{succ1} \left\{
\begin{array}{cl}
 (3) & \\
 (3) & \stackrel{1}{\rightsquigarrow} (3)^{3}\\
 (3) & \stackrel{2}{\rightsquigarrow} (3)^{2}\\
 (3) & \stackrel{3}{\rightsquigarrow} (\overline{3}) \,
\end{array}
\right.
\end{equation}


\subsection{Elimination of jumps and marked labels}\label{s2}

The successive step of our method consists into recursively
eliminating jumps from the extended succession rule (\ref{mark})
in order to obtain a finite succession rule which is equivalent to
the previous one. Once jumps have been eliminated we will deal
with marked labels.

\begin{proposition}\label{equiplus}
The succession rule:
\begin{equation} \label{succECO} \left\{
\begin{array}{cl}
 (a_1) & \\
 (a_1) & \rightsquigarrow (a_1 + a_2)(a_1)^{a_1 -1}\\
 (a_1 + a_2) & \rightsquigarrow (a_1 + a_2 + a_3)(a_1)^{a_1 + a_2 - 1}\\
  \vdots\\
 (\sum_{l=1}^{k-1}a_l) & \rightsquigarrow
 (\sum_{l=1}^{k}a_l)(a_1)^{(\sum_{l=1}^{k-1}a_l)-1}\\
 (\sum_{l=1}^{k}a_l) & \rightsquigarrow (\sum_{l=1}^{k}a_l)(a_1)^{(\sum_{l=1}^{k}a_l)-1}
\end{array}
\right.
\end{equation}
 is equivalent to the
recurrence relation $f_n = a_1 f_{n-1} + a_2 f_{n-2} + \dots + a_k
f_{n-k}$, $a_i \in \mathbb{Z}, 1 \leq i \leq k$, with default
initial conditions.
\end{proposition}

\textbf{Proof.} \quad Let $A_k(x)$ be the generating function of
the label $(\sum_{l=1}^{k}a_l)$ related to the succession rule
(\ref{succECO}). We have:
\begin{displaymath}
\begin{array}{l}
A_1(x)=1+(a_1-1)xA_1(x) + (a_1+a_2-1)xA_2(x)+ \dots +
(a_1+a_2+\dots+a_k-1)x A_k(x); \\\\
A_2(x)=x A_1(x); \\\\
A_3(x)=x A_2(x)= x^2A_1(x);\\
\hspace{10pt}\vdots\\
A_{k-1}(x)=x A_{k-2}(x)=x^{k-2}A_1(x); \\\\
A_k(x)=xA_{k-1}(x)+xA_k(x)= \frac{x^{k-1}}{1-x}A_1(x).
\end{array}
\end{displaymath}
Therefore,
\begin{displaymath}
A_1(x) = 1 + x(a_1-1)A_1(x)+ x^2(a_1 +a_2-1)A_1(x)+ \dots +
\frac{x^k}{1-x}(a_1+a_2+\dots+a_k-1)A_1(x),
\end{displaymath}
and we obtain the generating function
$A_1(x)=\frac{1-x}{1-a_1x-a_2x^2-\dots-a_kx^k} \ .$

At this point we can consider the generating function determined
by the succession rule (\ref{succECO}) as following:
$$A_1(x) + A_2(x) + \dots + A_{k-1}(x)+A_k(x)= A_1(x) +
xA_1(x) + \dots + x^{k-2}A_1(x)+\frac{x^{k-1}}{1-x}A_1(x)=$$
$$ =\frac{(1-x)+ x(1-x)+ \dots + x^{k-2}(1-x) +
x^{k-1}}{1-a_1x-a_2x^2-\dots-a_kx^k}=
\frac{1}{1-a_1x-a_2x^2-\dots-a_kx^k} \ .$$ \cvd

\begin{figure}[htb]
\begin{center}
\epsfig{file=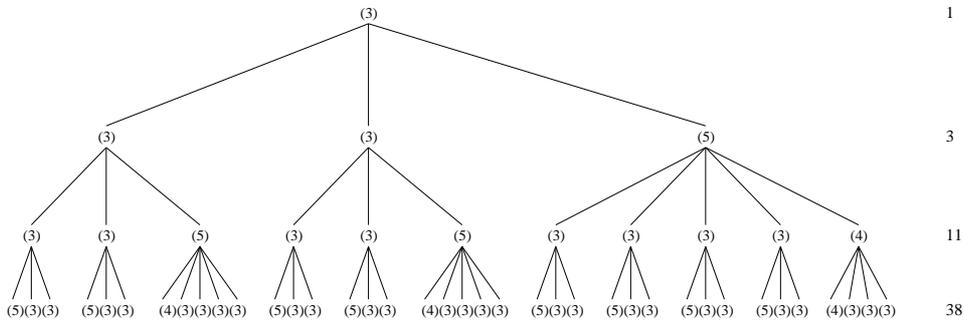,width=5in,clip=} \caption{\small{Four
levels of the generating tree associated with the succession rule
(\ref{succ2})} \label{ese2}}\vspace{-15pt}
\end{center}
\end{figure}

Following the previous statement, the extended succession rule
(\ref{succ1}) -- determined in the previous section --
can be translated into the following succession rule:
\begin{equation} \label{succ2} \left\{
\begin{array}{cl}
 (3) & \\
 (3) & {\rightsquigarrow} (5)(3)^{2}\\
 (5) & {\rightsquigarrow} (4)(3)^{4}\\
 (4) & {\rightsquigarrow} (4)(3)^{3}
\end{array}
\right.
\end{equation}
Figure \ref{ese2} shows some levels of the  generating tree
associated with (\ref{succ2}).

We observe that the previously obtained succession rule is an
ordinary finite succession rule, but it may happen that the value
of the label $(\sum_{l=1}^{i}a_l)$ is negative, for some $i$ with
$i \leq k$, then the succession rule (\ref{succECO}) contains
marked labels.

\begin{figure}[htb]
\begin{center}
\epsfig{file=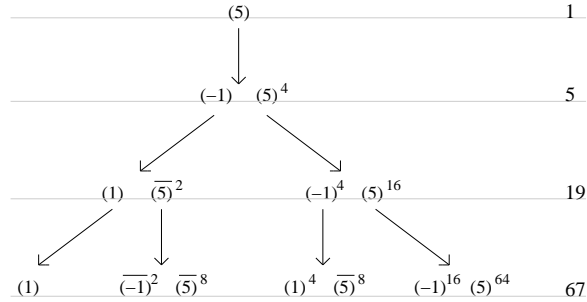,width=3in,clip=} \caption{\small{Compact
notation for the generating tree associated with the succession
rule (\ref{succ3})} \label{ese3}}\vspace{-15pt}
\end{center}
\end{figure}

For example, the recurrence relation
$f_n=5f_{n-1}-6f_{n-2}+2f_{n-3}$, with default initial conditions,
which defines the sequence 1,5,19,67,231,791,2703, \ldots,
(sequence A035344 in the The On-Line Encyclopedia of Integer
Sequences) is equivalent to the following succession rule:
\begin{equation} \label{succ3} \left\{
\begin{array}{cl}
 (5) & \\
 (5) & {\rightsquigarrow} (-1)(5)^{4}\\
 (-1) & {\rightsquigarrow} (1)(\overline{5})^{2}\\
 (1) & {\rightsquigarrow} (1)
\end{array}
\right.
\end{equation}
and Figure \ref{ese3} shows some levels of the associated
generating tree represented using a ``compact notation'', i.e., by
convention, the number of nodes at a given level $n$ is obtained
by means of the algebraic sum of the exponents of the labels lying
at level $n$.

\medskip

Therefore our next goal is to remove all possible marked labels
from the succession rule. We observe that in order to obtain this
goal, the recurrence relation $f_n = a_1 f_{n-1} + a_2 f_{n-2} +
\dots + a_k f_{n-k}$ with default initial conditions needs
$a_1>0$. We assume that this condition holds throughout the rest
of the present section.

\smallskip

In order to furnish a clearer description of our method, we start
considering the case $k=2$.

\begin{proposition}\label{p11}
The C-finite recurrence $f_n = a_1 f_{n-1} + a_2 f_{n-2}$, with
default initial conditions, and having $a_1>0$, is equivalent to
\begin{equation} \label{succECOneg2} \left\{
\begin{array}{cl}
 (a_1) & \\
 (a_1) & \rightsquigarrow (0)^{q_2}(r_2)(a_1)^{a_1 -(q_2+1)}\vspace{4pt}\\
 (r_2) & \rightsquigarrow \Big((0)^{q_2}(r_2)\Big)^{q_2}(0)^{q_2}(r_2)(a_1)^{r_2-(q_2+1)^2} \,
\end{array}
\right.
\end{equation}
where, by convention, the label $(0)$ does not produce any son,
and $q_2,r_2$ are defined as follows: \item{-} if $a_1+a_2 \leq 0$
then $q_2,r_2 >0$ such that $|a_1+a_2|=q_2a_1 - r_2$; \item{-}
otherwise $q_2=0$, $r_2=a_1+a_2$.
\end{proposition}


\textbf{Proof.} \quad We have to distinguish two cases: in the
first one $a_1+a_2 \leq 0$ and in the second one $a_1+a_2>0$.

If $a_1+a_2 \leq 0$, we have to prove that the generating tree
associated to the succession rule (\ref{succECOneg2}) is obtained
by performing some actions on the generating tree associated to
the extended succession rule (\ref{mark3}) which is obviously
equivalent to the recurrence $f_n = a_1 f_{n-1} + a_2 f_{n-2}$
having $a_1>0$ and $a_2<0$, with $f_0=1$ and $f_h=0$ for each
$h<0$.
\begin{equation} \label{mark3} \left\{
\begin{array}{cl}
 (a_1) & \\
 (a_1) & \stackrel{1}{\rightsquigarrow} (a_1)^{a_1}\\
 (a_1) & \stackrel{2}{\rightsquigarrow} (a_1)^{a_2}
\end{array}
\right.
\end{equation}

\begin{figure}[htb]
\begin{center}
\epsfig{file=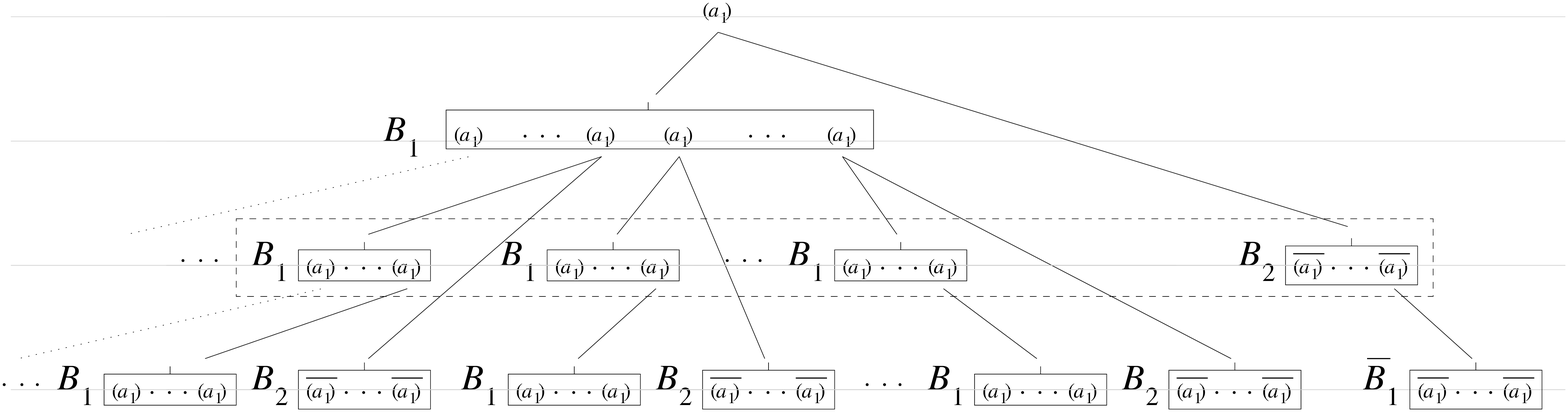,width=5in,clip=} \caption{\small{Step
1} \label{dimo1}}\vspace{-15pt}
\end{center}
\end{figure}

The proof consists in eliminating jumps and marked labels at each
level of the generating tree associated with succession rule
(\ref{mark3}), sketched in Figure \ref{dimo1}, by modifying the
structure of the generating tree, still maintaining $f_n$ nodes at
level $n$, for each $n$.

Let $(a_1)$ be a label at a given level $n$. We denote by $B_1$
the set of $a_1$ labels $(a_1)$ at level $n+1$ and by $B_2$ the
set of $a_2$ labels $(a_1)$ at level $n+2$, see Figure
\ref{dimo1}. We remark that $(a_1)^{a_2} =
\underbrace{\overline{(a_1)}\ldots\overline{(a_1)}}_{-a_2}.$

In order to eliminate both jumps and marked labels in $B_2$ at
level $2$ produced by the root $(a_1)$ at level $0$, we have to
consider the set of $a_1$ labels $(a_1)$ in $B_1$ at level 2
obtained by $(a_1)$ which lie at level $1$. At level 2, each label
$(a_1)$ in a given set $B_1$ kills one and only one marked label
$\overline{(a_1)}$ in $B_2$. At this point $|a_1+a_2|$ labels
$\overline{(a_1)}$ in $B_2$ always exist at level 2.

In order to eliminate such marked labels we have to consider more
than a single set $B_1$ of label $(a_1)$ at level 2. Let $q_2$ be
a sufficient number of sets $B_1$ at level 2 able to kill all the
labels $\overline{(a_1)}$ in $B_2$ at level 2. Therefore
$|a_1+a_2|=q_2 a_1 - r_2$ with $q_2, r_2 > 0$.

\begin{figure}[htb]
\begin{center}
\epsfig{file=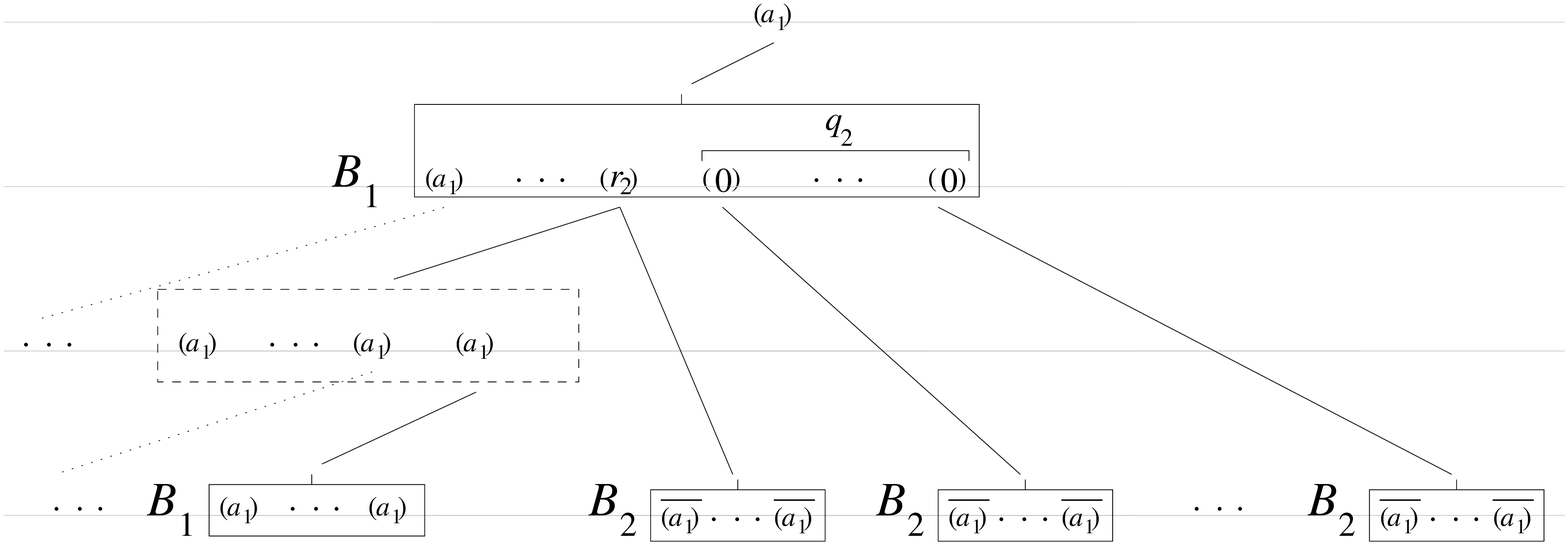,width=5in,clip=} \caption{\small{Step
2} \label{dimo2}}\vspace{-15pt}
\end{center}
\end{figure}

By setting $q_2$ labels $(a_1)$ at level 1 equal to $(0)$ and one
more label $(a_1)$ to $(r_2)$, we have the desired number of
labels $(a_1)$ at level 2. Note that the marked labels at level 2
are not generated and the labels $(a_1)$ at level 1 are revised in
order to have the right number of labels at level 2, see Figure
\ref{dimo2}.

Note that, when a label $(a_1)$ kills a marked label
$\overline{(a_1)}$ at a given level $n$, then the subtree, having
such label $(a_1)$ as its root, kills the subtree having
$\overline{(a_1)}$ as its root. So, at level 2 when a label
$(a_1)$ of $B_1$ kills a label $\overline{(a_1)}$ of $B_2$ then
the two subtrees having such labels as their roots are eliminated
too, see Figure \ref{dimo2}.

\begin{figure}[htb]
\begin{center}
\epsfig{file=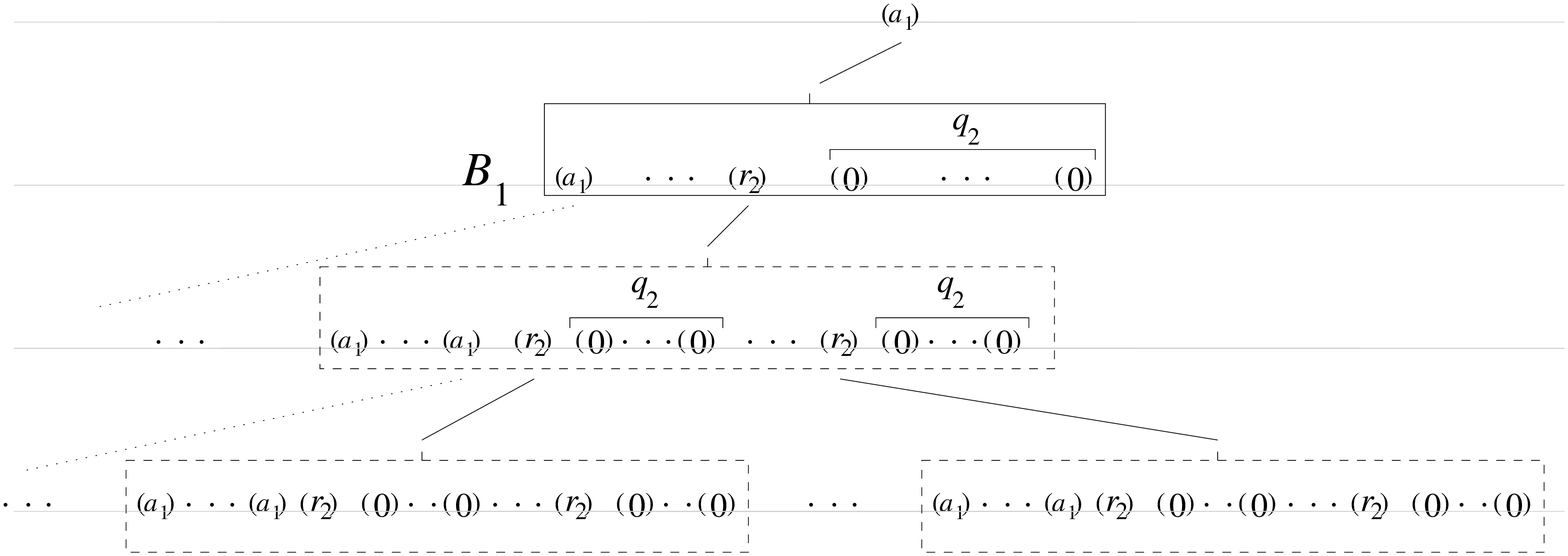,width=5in,clip=} \caption{\small{Step
3} \label{dimo3}}\vspace{-15pt}
\end{center}
\end{figure}

On the other hand, the $q_2+1$ sets $B_2$ at level 3 obtained by
the $q_2+1$ labels at level 1, once labelled with $(a_1)$ and now
having value $r_2,0,\dots,0$, respectively, are always present in
the tree, see Figure \ref{dimo2}. In order to eliminate such
undesired marked labels we can only set the production of $(r_2)$.
As a set $B_2$ at a given level is eliminated by using $q_2+1$
labels at previous level then $(r_2)$ must give
$(r_2)\underbrace{(0) \ldots (0)}_{q_2}$ exactly $q_2+1$ times.
This explains the first part of the production rule of the label
$(r_2)$ in rule (\ref{succECOneg2}). Since $(r_2)$ has $r_2$ sons
then the remaining $r_2 - (q_2+1)^2$ labels are set to be equal to
$(a_1)$ as in the previous case, see Figure \ref{dimo3}.

By the way, the modified $q_2+1$ labels having value
$r_2,0,\ldots,0$, respectively, at a given level $n$, produce the
labels $\Big((0)^{q_2}(r_2)\Big)^{q_2+1}(a_1)^{r_2-(q_2+1)^2}$ at
level $n+1$. Just as obtained for levels 1 and 2, the labels
$\Big((0)^{q_2}(r_2)\Big)^{q_2+1}$ automatically annihilate the
remaining $q_2+1$ sets $B_2$ of marked labels at level $n+2$, once
obtained by the modified $q_2+1$ labels at level $n$, see Figure
\ref{dimo3}.

Till now we have modified a portion $P$ of the total generating
tree in a way that it does not contain any marked label. Note
that, the remaining labels $(a_1)$ will be the roots of subtrees
which are all isomorphic to $P$.

The value $f_n$ defined by the tree associated to the extended
succession rule (\ref{mark3}), is given by the difference between
the number of non-marked and marked labels. The just described
algorithm modifies the number of generated non-marked labels and
sets to 0 the number of marked ones in a way that $f_n$ is
unchanged, for each $n$, so the succession rule
(\ref{succECOneg2}) is equivalent to the recurrence $f_n = a_1
f_{n-1} + a_2 f_{n-2}$.

In the case $a_1+a_2>0$ we have marked labels only if $a_2<0$. In
this case a single set $B_1$ is sufficient to kill all the marked
labels in $B_2$ at level 2. By the way, both in the case $a_2<0$
and $a_2>0$ we have that $q_2=0$ and $r_2=a_1+a_2$, and the
succession rule (\ref{succECOneg2}) has the same form of the rule
(\ref{succECO}) which is equivalent to the recurrence $f_n = a_1
f_{n-1} + a_2 f_{n-2}$ having $a_1>0$ and $a_2 \in \mathbb{Z}$,
with $f_0=1$ and $f_h=0$ for each $h<0$. \cvd

The statement of Proposition \ref{p11} can be naturally extended
to the general case $k>2$.

\begin{proposition}\label{pp}
The C-sequence $\{ f_n\} _n$ satisfying $f_n = a_1 f_{n-1} + a_2
f_{n-2} + \dots + a_k f_{n-k}$, with default initial conditions
and $a_1>0$ is equivalent to
\begin{equation} \label{succECOneg} \left\{
\begin{array}{cl}
 (a_1) & \\
 (a_1) & \rightsquigarrow (0)^{q_2}(r_2)(a_1)^{a_1 -(q_2+1)} \vspace{4pt}\\
 (r_2) & \rightsquigarrow \Big((0)^{q_2}(r_2)\Big)^{q_2}(0)^{q_3}(r_3)(a_1)^{r_2-(q_2(q_2+1)+q_3+1)}\\
  \vdots\\
 (r_i) & \rightsquigarrow \Big((0)^{q_2}(r_2)\Big)^{q_i}(0)^{q_{i+1}}(r_{i+1})(a_1)^{r_i-(q_i(q_2+1)+q_{i+1}+1)}
 \ \ , \ \ 3 \leq i \leq k-1 \\
 \vdots\\
 (r_k) & \rightsquigarrow \Big((0)^{q_2}(r_2)\Big)^{q_k}(0)^{q_k}(r_k)(a_1)^{r_k-(q_k(q_2+1)+q_k+1)}
\end{array}
\right.
\end{equation}
where the parameters $q_i$ and $r_i$, with $2 \leq i \leq k$, can
be determined in the following way: \item{-} if $\sum_{l=1}^{i}
a_l \leq 0$ then $q_i, r_i > 0$ such that $|\sum_{l=1}^{i}
a_l|=q_i a_1 - r_i$, \item{-} otherwise $q_i=0$ and
$r_i=\sum_{l=1}^{i} a_l$.
\end{proposition}

The proof of the Proposition \ref{pp} is quite similar to the
proof of Proposition \ref{p11}. It has the same level of
difficulty but it is more cumbersome, so it is omitted for
brevity.

Using Proposition \ref{pp}, we can translate the previously
considered recurrence relation $f_n=5f_{n-1}-6f_{n-2}+2f_{n-3}$,
with default initial conditions, into the following ordinary
succession rule:
\begin{equation} \label{succ4} \left\{
\begin{array}{cl}
 (5) & \\
 (5) & {\rightsquigarrow} (0)(4)(5)^{3}\\
 (4) & {\rightsquigarrow} (0)(4)(1)(5)\\
 (1) & {\rightsquigarrow} (1)
\end{array}
\right.
\end{equation}

being $q_2=1$, $r_2=4$, $q_3=0$ and $r_3=1$.


\subsection{Positivity condition}\label{s3}

The statement of Proposition  \ref{pp} is indeed a tool to
translate C-recurrences into finite succession rules. However this
property turns out to be effectively applicable only when the
labels of the succession rule are all positive, and the reader can
easily observe that Proposition  \ref{pp} does not give us an
instrument to test whether this happens or not.

In particular, if the labels of the succession rule are all
positive then the terms of the C-sequence are all positive. It is
then interesting to relate our problem with the so called {\em
positivity problem}, which we have already mentioned in the
Introduction.

\medskip

\noindent {\bf Positivity Problem:} given a C-finite sequence
$\{f_n \}_{n\geq 0} $, establish if all its terms are positive.

\medskip

We recall that the problem was originally proposed as an open
problem in \cite{BM}, and then re-presented in \cite{SS} (Theorems
12.1-12.2, pages 73-74), but no general solution has been found
yet.

Moreover, the positivity problem can be solved for a large class
of C-finite sequences, precisely for $\Bbb{N}$-rational sequences.
We recall that the class of $\Bbb{N}$-rational series is precisely
the class of the generating functions of regular languages, and
that Soittola's Theorem \cite{S}  states  that the problem of
establishing whether a rational generating function is
$\Bbb{N}$-rational is decidable.

Let us start examining the case of C-recurrences of degree $2$.
So, let $f_n = a_1 f_{n-1} + a_2 f_{n-2}$ be a recurrence
relation, with $a_1>0$ and $a_2 \in \mathbb{Z}$.

By referring to the succession rule (\ref{succECOneg2}), precisely
to the case $a_1+a_2 \leq 0$, we observe that the succession rule
equivalent to the recurrence relation is an ordinary rule (i.e.,
it has all positive labels) if and only if the following condition
is verified:
\begin{equation}\label{su2}\left\{
\begin{array}{l}
a_1 -(q_2+1) \geq 0 \\
r_2 -(q_2+1)^2 \geq 0
\end{array}
\right.
\end{equation}

As $r_2=q_2a_1 -|a_1+a_2|=q_2a_1 + a_1 + a_2$ then $r_2 -(q_2+1)^2
\geq 0$ means ${q_2}^2 +(2-a_1)q_2 +1 -a_1-a_2 \leq 0$. This
inequality has solution if and only if ${a_1}^2+4a_2 \geq 0$, and
this is clearly a necessary and sufficient  condition to ensure
the positivity of all the terms of $f_n$ \cite{BR} .

Let us now consider a generic C-recurrence of degree $k$. Using a
similar reasoning, and following  Proposition  \ref{pp} we can
prove:

\begin{corollary}\label{result}
Let  us consider the recurrence relation $f_n = a_1 f_{n-1} + a_2
f_{n-2} + \dots + a_k f_{n-k}$ having $a_1>0$ and $a_i \in
\mathbb{Z}$, $2 \leq i \leq k$, with $f_0=1$ and $f_{h}=0$ for
each $h<0$. If
\begin{equation}\label{posity}\left\{
\begin{array}{l}
a_1 -(q_2+1) \geq 0 \\
r_2-(q_2(q_2+1)+q_3+1) \geq 0\\
\vdots\\
r_i-(q_i(q_2+1)+q_{i+1}+1) \geq 0 \ \ , \ \ 3 \leq i \leq k-1 \\
\vdots\\
r_k-(q_k(q_2+1)+q_k+1) \geq 0
\end{array}
\right.
\end{equation}
then $f_n > 0$ for all $n$.
\end{corollary}

As $r_i=\sum_{l=1}^{i} a_l +  q_i a_1$, $2 \leq i \leq k$, then
the system (\ref{posity}) can be rewritten as
\begin{equation}\label{posity2}\left\{
\begin{array}{l}
a_1 -(q_2+1) \geq 0 \\
\vdots\\
\sum_{l=1}^{i} a_l +  q_i a_1-(q_i(q_2+1)+q_{i+1}+1) \geq 0 \ \ , \ \ 2 \leq i \leq k-1 \\
\vdots\\
\sum_{l=1}^{k} a_l +  q_k a_1-(q_k(q_2+1)+q_k+1) \geq 0 .
\end{array}
\right.
\end{equation}

As previously mentioned, condition (\ref{posity2}) ensures that
all the labels of the succession rules equivalent to the given
C-recurrence relation are positive, hence all the terms $f_n$ are
positive. Thus it can be viewed as a sufficient condition to test
the positivity of a given C-recurrence relation.

\medskip

Unfortunately, this is not a necessary condition to test
positivity, then there are cases of positive C-sequences for which
our method fails to prove positivity. A simple example is given by
any positive non $\mathbb{N}$-rational C-sequence. The reader can
find an instance of such sequences in \cite{K}. It would be more
interesting to give an example of a $\mathbb{N}$-rational
C-sequence for which our method is not able to prove positivity,
but we have not been able to find any such example.

\medskip

Clearly, any C-sequence satisfying the positivity condition has a
$\mathbb{N}$-rational generating function (in fact, any finite
succession rule may be regarded as a finite state automaton), thus
our method can be suitably used to test the
$\mathbb{N}$-rationality of a sequence. Though it is not our
intention to deepen the computational complexity of our test, we
remark that, despite the methods presented in \cite{BR,K,P}, our
method does not deal with calculating polynomial roots.

\medskip

 In order to give an idea
of the computational cost to solve the system (\ref{posity2}) we
consider the worst case that is when $\sum_{l=1}^{i} a_l \leq 0$,
$2 \leq i \leq k$, and the system itself has no solution.

In this case all the possible values for each $q_i$, $2 \leq i
\leq k$, must be checked in order to conclude that the system
(\ref{posity2}) does not admit any solution.

As $q_2$ can range in the close set $[1,a_1-1]$ and $q_{i+1}$ in
$[1, \sum_{l=1}^{i} a_l +  q_i a_1-(q_i(q_2+1)-1]$ then we have
$$1 + (a_1 - 1) \prod_{i=2}^{k-1}\big{(}\sum_{l=1}^{i} a_l +
a_1 q_i - q_i (q_2+1)-1 \big{)}$$

\noindent where the first 1 accounts the check to verify
$\sum_{l=1}^{k} a_l + q_k a_1-(q_k(q_2+1)+q_k+1) \geq 0$.

An average complexity study of our test is a further development.
Anyway, the referred experimental results give a sufficiently
short computational time to test condition (\ref{posity2}).

\subsection{Generic initial conditions}\label{s4}

Now it is possible to use the statement of Proposition \ref{pp} to
treat the case of C-recurrence relations with generic initial
conditions. The following result is obtained by simply adapting
the productions of the labels in the first levels of the
generating tree to the given initial conditions, then using the
productions of Proposition \ref{pp}. So, we have two sets of
productions: the ones stating the initial conditions, and the
remaining ones defining all the other levels.

\begin{proposition}\label{pq}
Let us consider the C-finite recurrence relation $f_n = a_1
f_{n-1} + a_2 f_{n-2} + \dots + a_k f_{n-k}$, $a_i \in
\mathbb{Z}$, $1 \leq i \leq k$, and let us assume that the initial
conditions are $f_0=1$ and $f_i=h_i$, with $h_i \in \mathbb{Z}$,
$1 \leq i < k$, then it can be translated into the following
extended succession rule:
\begin{equation} \label{markci} \left\{
\begin{array}{cl}
 (h_1) & \\
 (h_1) & \stackrel{1}{\rightsquigarrow} (a_1)^{h_1}\\
 \vdots\\
 (h_1) & \stackrel{i}{\rightsquigarrow} (a_1)^{h_i - \sum_{j=1}^{i-1}h_j a_{i-j}} \ \ , \ \ 1 < i < k \\
 \vdots\\
 (h_1) & \stackrel{k}{\rightsquigarrow} (a_1)^{a_k}\\\\
 (a_1) & \stackrel{1}{\rightsquigarrow} (a_1)^{a_1}\\
 (a_1) & \stackrel{2}{\rightsquigarrow} (a_1)^{a_2}\\
 \vdots\\
 (a_1) & \stackrel{k}{\rightsquigarrow} (a_1)^{a_k}
\end{array}
\right.
\end{equation}
\end{proposition}

\begin{figure}[htb]
\begin{center}
\epsfig{file=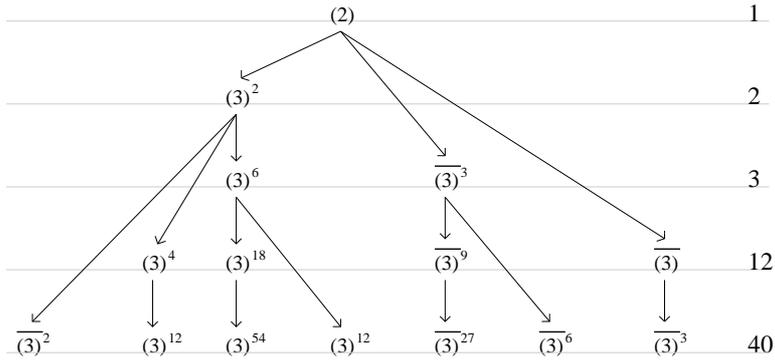,width=4in,clip=}
\caption{\small{Compact notation for the generating tree
associated with the succession rule (\ref{mci})}
\label{esci}}\vspace{-15pt}
\end{center}
\end{figure}

For example, the recurrence relation $f_n=
3f_{n-1}+2f_{n-2}-f_{n-3}$ with $f_0=1$, $f_1=2$ and $f_2=3$,
which defines the sequence $1,2,3,12,40,141,491,1715, \ldots$, is
equivalent to the following extended succession rule:
\begin{equation} \label{mci} \left\{
\begin{array}{cl}
 (2) & \\
 (2) & \stackrel{1}{\rightsquigarrow} (3)^{2}\\
 (2) & \stackrel{2}{\rightsquigarrow} (\overline{3})^{3}\\
 (2) & \stackrel{3}{\rightsquigarrow} (\overline{3})\\\\
 (3) & \stackrel{1}{\rightsquigarrow} (3)^{3}\\
 (3) & \stackrel{2}{\rightsquigarrow} (3)^{2}\\
 (3) & \stackrel{3}{\rightsquigarrow} (\overline{3})
\end{array}
\right.
\end{equation}
and Figure \ref{esci} shows some levels of the generating tree
associated to it.

Following the described method in Section \ref{s2} to eliminate
jumps and marked labels, we can translate the extended succession
rule (\ref{mci}) into the ordinary succession rule (\ref{oci}),
where the labels $(3)$, $(3)_1$ and $(3)_2$ are different labels
with different productions.
\begin{equation} \label{oci} \left\{
\begin{array}{ll}
 (2) & \\
 (2) & {\rightsquigarrow} \ (0) (3)_1\\
 (3)_1 & {\rightsquigarrow} \ (6)(3)^{2}\\
 (6) & {\rightsquigarrow} \  (3)^{5} (3)_2\\
 (3)_2 & {\rightsquigarrow} \ (4)(3)^{2}\\\\
 (3) & {\rightsquigarrow} \ (5)(3)^{2}\\
 (5) & {\rightsquigarrow} \ (4)(3)^{4}\\
 (4) & {\rightsquigarrow} \ (4)(3)^{3}
\end{array}
\right.
\end{equation}

\section{Concluding remarks}

In this paper we have presented a general method to translate a
given C-finite recurrence relation into an ordinary succession
rule and we have proposed a sufficient condition for testing the
positivity of a given C-finite sequence.

A further development could take into consideration the average
complexity necessary to prove the positivity of a given C-finite
sequence.

Afterwards, it should be interesting to develop the study
concerning the C-recurrence relations with generic initial
conditions in order to examine in depth the potentiality of our
method.

\smallskip

Finally, we would like to show that some of our ideas
can be applied to the case of holonomic integer sequences, i.e.,
those satisfying a linear recurrence relation with polynomial
coefficients.

Just to have a simple example, let us consider the {\em
involutions} of $n$, enumerated by the sequence $\{f_n\}$ defined
by the holonomic recurrence relation
\begin{equation}\label{i1}
f_n=f_{n-1}+(n-1)f_{n-2},
\end{equation}
with $f_0=1$, $f_1=1$ (sequence A000085 in the The On-Line
Encyclopedia of Integer Sequences).

We easily observe that, using the same argument of Proposition
\ref{equiplus}, we can translate the recurrence relation
(\ref{i1}) into an infinite succession rule (possibly having
marked labels and jumps), where now we adopt the convention that a
generic label $(k)$ is placed at the level $k$ of the generating
tree:
\begin{equation}\label{i2}
 \left\{
\begin{array}{cl}
  (0) & \\
 (k)  & \stackrel{1}{\rightsquigarrow} (k+1)\\
 (k)  & \stackrel{2}{\rightsquigarrow} (k+2)^{k+1} \\
 \end{array}
\right.
\end{equation}
The successive step is to find a way how to convert such a rule
into an ordinary succession rule. Referring to (\ref{i2}), this
can be done by eliminating  ``by hand'' marked labels and jumps,
then re-writing the (ordinary) rule as follows:
\begin{equation}\label{i3}
 \left\{
\begin{array}{cl}
 (1)  & \\
 (k)  & {\rightsquigarrow} (k-1)^{k-1}(k+1)\\
 \end{array}
\right.
\end{equation}

We believe that such a method should be formalized in some further
work, and then applied to automatically convert the obtained rule
into an ordinary succession rule.

Moreover, from this method, we could also develop a more general
criterion for proving the positivity of an holonomic sequence.

\end{document}